\documentclass[pdftex]{article} 
\usepackage{amsmath} 
\usepackage{geometry} 
\usepackage{rotating} 
\usepackage{graphicx} 

\title{Accelerating the pace of discovery by changing the peer review algorithm}

\author{Stefano Allesina$^{1,2}$\\ \small $^1$ Department of Ecology and
  Evolution and Computation Institute,\\
  \small University of Chicago, 1101
  E. 57th Street, Chicago, IL 60637\\ \small $^2$ National Center for
  Ecological Analysis and Synthesis,\\ \small 735 State Street, Santa Barbara, CA
  93101, USA}
\begin{document}
\maketitle
\begin{abstract}
  The number of scientific publications is constantly rising, increasing
  the strain on the review process. The number of submissions is
  actually higher, as each manuscript is often reviewed several times
  before publication\cite{aarssen2008bang}. To face the deluge of
  submissions, top journals reject a considerable fraction of
  manuscripts without review\cite{NatureRej}, potentially declining
  manuscripts with merit. The situation is frustrating for
  authors\cite{smit2006peer}, reviewers and
  editors\cite{schroter2004effects,smith2006peer,hochberg2009tragedy,mcpeek2009golden}
  alike. Recently, several editors wrote about the ``tragedy of the
  reviewer commons''\cite{hochberg2009tragedy}, advocating for urgent
  corrections to the system. Almost every scientist has ideas on how to
  improve the
  system\cite{cintas2004confidential,list2006reviewers,hauser2007incentive,CARMA,campanario1997journal,ImprovingMSsel,Marketpeer},
  but it is very difficult, if not impossible, to perform experiments to
  test which measures would be most effective. Surprisingly, relatively
  few attempts have been made to model peer
  review\cite{neff2006peer,bentley2009game}. Here I implement a
  simulation framework in which ideas on peer review can be
  quantitatively tested. I incorporate authors, reviewers, manuscripts
  and journals into an agent-based model and a peer review system
  emerges from their interactions. As a proof-of-concept, I contrast an
  implementation of the current system, in which authors decide the
  journal for their submissions, with a system in which journals bid on
  manuscripts for publication. I show that, all other things being
  equal, this latter system solves most of the problems currently
  associated with the peer review process. Manuscripts' evaluation is
  faster, authors publish more and in better journals, and reviewers'
  effort is optimally utilized. However, more work is required from
  editors. This modeling framework can be used to test other solutions
  for peer review, leading the way for an improvement of how science is
  disseminated.
\end{abstract}
\noindent
Typically, a reviewer is asked to evaluate a manuscript according to
three main characteristics: i) Topic ($T$): Is the topic of the
manuscript in line with the journal?  ii) Technical Quality ($Q$): Does
the work use the best techniques? Are references adequate?  Is the
statistical treatment of the data correct?  iii) Novelty ($N$): Is the
work groundbreaking? Does it convey new ideas and methods?  Accordingly,
in the model each author and journal is defined by three beta
probability distribution functions, one for each characteristic ($T$,
$Q$ and $N$, Methods, Fig. 1-i). A manuscript is simply a sample from
the three distributions for the author ($t,\ q,\ n$ - Fig. 1-ii). Once
an author has produced a manuscript, she has to decide where to submit
it. In the current setting ($CS$), an author is encouraged to choose
optimally the journal so that the impact of the publication on other
scientists (number of readers, citations) is maximized and so is the
impact on her career (grants, tenure). Let us assume that the impact of
a journal is measurable (impact factor, eigenfactor) and that the author
is able to determine with precision the probability of acceptance of her
manuscript in a given journal. Then the author can optimally choose the
journal by maximizing the product of acceptance probability and impact
(Fig. 1-iii). Note that this procedure postulates that authors have a
cost associated to the submission (e.g. time spent editing,
reputation). If there is no cost, authors will tend to submit always to
the highest impact journals.

In the $CS$, once the manuscript is submitted to a journal, the editor
assigns it to three referees. The referees are sampled at random from
those whose expertise is best suited for the manuscript (Methods). The
referees then provide the editor with an estimate of the three values
characterizing the manuscript. The more familiar the reviewers are with
the topic, the more accurate their estimate is (Methods). The author
revises the manuscript (increasing its quality and novelty, Methods) and
the review process is repeated (Fig. 1-iv). Finally, the editor averages
the $t$, $q$, and $n$ values provided by the referees and uses this
averaged estimates to decide the fate of the manuscript. This is done in
the model by computing an acceptance probability ($p_a$). This
probability is the product of the cumulative distribution functions of
the journal for the estimate $q$ and $n$, times a measure based on $t$
that determines the interest of the journal in the topic of the
manuscript (Methods, Fig. 1-v). The editor then accepts the manuscript
with probability $p_a$ and rejects it with probability $1-p_a$. In the
$CS$, if a manuscript is rejected then the author will submit it to the
next best journal, unless it has been rejected 5 times (Methods), in
which case the author abandons the manuscript.

In the alternative setting ($AS$, Methods), when an author produces a
new manuscript, she will submit it to a first pool of manuscripts
(e.g. a preprint archive). However, to be able to submit one manuscript
the author must choose three manuscripts already in the pool for
review. Therefore, more productive authors are also the more active
reviewers. Once a manuscript in the first pool accrues three reviews, it
is revised (increase in quality and novelty), and the reviewers are
asked for a second evaluation. Then, the manuscript is moved to a second
pool (ripe manuscripts). Every month, the editors of the journals
evaluate the ripe articles. If an editor wants a manuscript for her
journal, she will bid on it. At the end of the month, authors receive
all the bids for their manuscripts in the second pool. In the case of
more than one journal bidding on her manuscript, the author will choose
that with the highest impact. If no journals bid on a manuscript, the
author abandons it.

%Simulations:
I implemented both the $CS$ and the $AS$ using 50 journals and 500
authors. To produce a fair comparison, the same authors and journals are
used in the two simulations. I ran the two models for 10 years and
tracked all manuscripts produced in this period (Methods).

%Results (using Nov 09): 
%Number of papers accepted, rejected submitted reviewed dropped.

A first coarse comparison can be done among the number of papers
published. In the $CS$ the 500 authors produced 14936 unique manuscripts
in 10 years. Of these, 9526 have been accepted for publication (63.8\%),
while 5410 have been abandoned after having been rejected by 5
journals. In the $AS$, instead, 14683 of the 14984 manuscripts found
their way into publication (98\%), while 301 were abandoned. Also the
number of reviews is extremely different. In the $CS$, on average 10.02
reviews per manuscript have been performed (the mean drops to 7.19
considering published manuscripts only), while in the $AS$ each
manuscript receives exactly 3 reviews. 92\% of the authors published
more papers in the $AS$ (Fig. 2-i), and all the authors accrued more
impact (computed summing the impact of the journal for each of the
author's articles). Surprisingly, the average impact for each author was
also higher in the $AS$ in all cases (Fig. 2-ii).  Also, the $AS$ system
is much faster: the average time to publication of the published papers
is 21.3 months in the $CS$ (1st Qu.:11, 3rd Qu.:30), but only 8.65 in
the $AS$ (1st Qu.:7, 3rd Qu.:10) (Fig. 2-iii).  From these results, it
seems that authors should prefer the $AS$: their articles are published
faster, fewer are abandoned, they are asked to perform less reviews and
they publish in higher-impact journals. Is this the better system for
journals as well?  The system requires a much greater load of work to
the editors: every month they have to go through all the ripe papers,
and this is likely to be a cost for the journals. On the other hand,
this cost in the $CS$ is payed by the reviewers, that are typically not
compensated. 58\% of the journals published more articles in the $AS$,
with an interesting trend: top journals (high impact) published more in
the $AS$, while lower-ranked journals published more in the $CS$
(Fig. 2-iv). If we measure the average merit of manuscripts as the
product of their $q$ and $n$ values, we find that merit is higher for
published papers in the $CS$ (mean: 0.378) and lower for $AS$
(0.323). This is due to two effects: a) in the $CS$ a larger fraction of
manuscripts is abandoned (their mean merit being 0.256), while in the
$AS$ only extremely scarce articles are abandoned (mean merit 0.056),
and b) the number of revision rounds in the $CS$ is much higher, and
each revision improves the merit of the manuscript (Methods).

In the $CS$, abandoned manuscripts can be reduced by allowing the
authors to submit to a greater number of journals before abandoning
manuscript. I repeated the analysis allowing up to 10 submissions for
each manuscript. This lead to the publication of 81.7\% of the
manuscripts (instead of 65\%). However, this also negatively impacted
the publication time (mean 32.53 months instead of 21.3) and the number
of reviews (mean 14.20 instead of 10.02). This is a trade-off that
cannot be avoided in the $CS$.

This exercise represents a highly idealized peer-review process. All the
reviews are useful, there is no bias, and each revision increases the
overall quality of the manuscript; authors always revise their
manuscript according to the referees' comments. Moreover, authors can
estimate exactly a) their manuscript value $(t\,,q\,,n)$ and b) the
probability of having the manuscript accepted in any journal. Note also
that if the cost associated with a submission is very small, then
authors are practically encouraged to always submit to top journals,
regardless the probability of acceptance, rendering the system even more
inefficient. The $CS$ can therefore be seen as a best case scenario for
the current system of peer review. Note that in the $AS$ submission does
not depend on the cost associated to it, as authors can not decide where
to submit their manuscript.

Results suggest that a much more efficient way of reviewing manuscripts
can be implemented: publication is faster, reviews are efficiently
distributed and proportional to productivity, authors publish more and
in better journals, better journals publish more articles. Moreover,
there could be other ``social'' advantages to the $AS$.  Recently,
several journals started providing ``Open Access'' (OA) content.
Typically, authors have to pay a fee to have their articles published as
OA. If the $AS$ were to be implemented, OA would be free and guaranteed,
as anybody could access the manuscripts in the two pools (as it happens
now, for example, for the {\em arXiv.org}). Scientists in developing
countries and poor institutions that can not afford subscription costs
would be on the cutting edge of science as their colleagues in more
fortunate situations. In fact, they could access not only articles that
have already been sanctioned by journals, but also the manuscripts in
the pools, i.e. what will be published in the future.  One could imagine
that a scientist could read published articles in her discipline, ``ripe
manuscripts'' in her sub-field and even submitted manuscripts on
specific topics. A parallel can be drawn with the release of open source
software: normal users will install ``stable'' versions, but those who
want to experience new features can install ``unstable''
releases. Another advantage for the authors publishing in the $AS$ is
that their work becomes visible as soon as the manuscript is submitted,
speeding up the process of accruing citations and settling eventual
priority disputes.

Finally, the $AS$ already embeds some of the changes that have been
proposed for the current peer review system\cite{campanario1997journal},
e.g. the creation of a database of reviews common to all
journals\cite{ImprovingMSsel}, and the idea of a ``Community-based
Assessment of Review Material''\cite{CARMA}. Both are practically
embedded in the $AS$. 

This alternative system, however, could give rise to different
concerns. For example, forcing the authors to review manuscripts could
be counter-productive when there are no suitable manuscripts in the
first pool. Also, manuscripts of less interest could not receive reviews
for a very long time. Finally, nothing prevents the abuse of the system,
where authors in cliques exchange positive reviews. Editors could
request additional ``traditional'' reviews for manuscripts whose
reviewers seem to be in conflict of interests, but this would slow down
the system.

However, thanks to the transparency of the system, most of these
problems could be easily solved: studying the authorship network editors
could find automatically the ``distance'' of the referees from the
authors, weighting their comments accordingly. Also, a rating system for
referees could be put in place so that authors and editors could assign
a rating to each review. Reviewers with outstanding ratings could be
good candidates for editorial work, providing an incentive to perform
balanced and constructive reviews.

There is an infinite number of ways to implement a peer review system,
and this work is a first attempt to tackle the evaluation of
alternatives in a quantitative way. I have examined only two options,
but with further developments the model can accommodate any variation on
the theme. For example, the simulations could be made more realistic
embedding multi-authored articles, editorial rejections, including a
reputation trait to the authors, allowing referees to decline
invitations to review.

%For example, one could imagine a system in which reviewers are
%paid by the authors to perform their reviews, so that authors are forced
%to decide for cost-effective submissions. Also, the simulations could be
%made more realistic embedding multi-authored articles, editorial
%rejections, including a reputation trait to the authors, allowing
%referees to decline invitations to review. 

%To make the test of this improvements and other alternatives easier, the
%code is available as open-source.
 
\section*{Methods}
  \noindent {\bf Author.} Each author is defined by three beta
  distributions, $T$, $Q$ and $N$. Each distribution $X$ requires two
  parameters: $\alpha_X$ and $\beta_X$. If both parameters are equal to
  1, the distribution is uniform $U[0,1]$. The beta distribution is
  quite flexible, so that various shapes can emerge.  For all the
  simulations presented here, I created 500 authors according to three
  types. 50 were ``broad interests'' authors whose parameters were
  extracted at random from uniform distributions: $\alpha_T$, $\beta_T$
  from $U[1,5]$, $\alpha_Q$ and $\alpha_N$ from $U[50,100]$, $\beta_Q$
  and $\beta_N$ from $U[5,10]$. These authors work on a broad range of
  topics with very high technical quality and novelty. I then added 150
  ``specialist'' authors: $\alpha_T$, $\beta_T$ from $U[10,100]$,
  $\alpha_Q$ and $\alpha_N$ from $U[5,10]$, $\beta_Q$ and $\beta_N$ from
  $U[1,5]$. The remaining 300 authors were assigned values using
  $U[1,10]$ for $\alpha_T$, $\beta_T$, $\alpha_Q$ and $\alpha_N$, while
  using $U[5,10]$ for $\beta_Q$, $\beta_N$.

  All the authors have 0.25 probability of producing a manuscript at
  each month. All manuscripts are single-authored. In the $CS$, authors
  re-submit a manuscript until it has been rejected 5 times, in which
  case they abandon the manuscript as unpublishable.

  \noindent {\bf Journal.} Also journals are defined by three beta
  distributions. I created a list of 50 journals using the following
  parameters: 5 ``broad interest'' journals ($\alpha_T$ and $\beta_T$
  from $U[1,5]$, $\alpha_Q$ and $\alpha_N$ from $U[50,100]$, $\beta_Q$
  and $\beta_N$ from $U[5,10]$); 15 ``specialist'' journals (
  $\alpha_T$, $\beta_T$ from $U[10,100]$, $\alpha_Q$ and $\alpha_N$ from
  $U[5,10]$, $\beta_Q$ and $\beta_N$ from $U[1,5]$); 30 ``normal''
  journals ($\alpha_T$, $\beta_T$, $\alpha_Q$, $\alpha_N$ from
  $U[1,10]$, $\beta_Q$, $\beta_N$ from $U[5,10]$).
 
  Journal impact depends on its distributions. It is computed as:
  \begin{equation}
    I=\frac{\alpha_Q}{\alpha_Q+\beta_Q}
    \frac{\alpha_N}{\alpha_N+\beta_N} \frac{1}
    {z(\alpha_T,\beta_T,\alpha_T/(\alpha_T+\beta_T),0.1)}
  \end{equation}
  where $B(\alpha,\beta,x)$ is a beta distribution and
  $z(\alpha_T,\beta_T, x, 0.1)$ is a measure of the density of a beta
  distribution around $x$. Note that in this case $x$ is simply the mean
  of the beta distribution.
  If $x+0.1 \leq 1$ and $x-0.1 \geq 0$ then $z$ is simply obtained by
  integrating the probability density function:
  \begin{equation}
    z(\alpha_T,\beta_T, x, 0.1) = \int^{x+0.1}_{x-0.1}
    B(\alpha_T,\beta_T,t) dt \; \; \; \; \; \; \;0.1 \leq x \leq 0.9
  \end{equation}
  However, in order to ensure that the computation of the impact is fair
  for journals whose average $T$ is close to 0 or 1, the following has
  to be implemented:
  \begin{equation}
    z(\alpha_T,\beta_T, x, 0.1) = \int^{1}_{x-0.1} B(\alpha_T,\beta_T,t)
    dt + \int^{1-x+0.1}_{0} B(\alpha_T,\beta_T,t) dt \; \; \; \; \; \;
    \; x \geq 0.9
  \end{equation}
  \begin{equation}
    z(\alpha_T,\beta_T, x, 0.1) = \int^{x+0.1}_{0} B(\alpha_T,\beta_T,t)
    dt + \int^{1}_{0.9+x} B(\alpha_T,\beta_T,t) dt \; \; \; \; \; \; \;
    x \leq 0.1
  \end{equation}

  This definition of $z(\alpha_T,\beta_T, x, 0.1)$ ensures that journal
  with mean topic close to 0 or 1 are not penalized. Therefore, the
  impact of a journal is simply the product of the expected value for
  $Q$ and $N$ (first two terms) divided by the degree of specialization
  of the journal: if most of the density is concentrated around the mean
  (highly specialized journals), then the impact will be lower, if
  instead the density around the mean is low (the minimum is 0.2 for a
  uniform distribution, i.e. a completely generalist journal) then the
  journal will have higher impact.  
  %The theoretical limit for the  impact is 5.

   \noindent {\bf CS: submission.} In the $CS$ authors maximize the
   expected impact of their submission. If a journal $X$ has impact
   $I_X$ and the probability of acceptance in the journal (see below)
   for an article whose values are $(t,q,n)$ is $p_X$ then the ``score''
   of the journal is $p_X I_X$. Authors rank the journals according to
   the score for their manuscript and submit to the highest ranked
   first. The submission happens 1 month after the manuscript is
   created.
  
   \noindent {\bf CS: reviewers' choice.} Journals choose the reviewers
   for a manuscript according to their expertise. The score of a
   reviewer $X$, whose $T$ distribution is defined by $\alpha_T^X$ and
   $\beta_T^X$ for a manuscript whose topic is $t$ is
   $1/z(\alpha_T^X,\beta_T^X,t,0.1)$ where $z$ is the function
   introduced above. Reviewers who are more familiar with the topic of
   the manuscript will therefore receive an higher score. Reviewers are
   sorted according to their score (descending order) and three
   reviewers are sampled from the top 20.
   
   \noindent {\bf CS: reviews.} Every month each author completes each
   of the assigned reviews with probability 0.5. The review process is
   simply an estimate of the ``true'' values of a manuscript
   $(t,q,n)$. The estimate has an error that is inversely proportional
   to the familiarity of the reviewer with the topic of the
   manuscript. We define $\delta$ as $\delta = \left(1-{\mit
   z}\left(\alpha_T,\beta_T, t,0.1 \right) \right)/2$ where the
   $\alpha_T$and $\beta_T$ are the parameters for the reviewer, $t$ is
   the real topic value for the manuscript to be reviewed and $z$ is the
   function introduced above to compute the density of a beta
   distribution around a given value ($t$). The estimate $t_1$ of the
   topic for the first reviewer is extracted from the uniform
   distribution $U[max(t-\delta,0), min(t+\delta,1)]$. In the same way
   $q_1$ is extracted from $U[max(q-\delta,0), min(q+\delta,1)]$, and
   $n_1$ from $U[max(n-\delta,0), min(n+\delta,1)]$. The three reviewers
   follow the same procedure, and finally the editor estimates are
   obtained averaging the three estimates for the reviewer:
   $t_e=(t_1+t_2+t_3)/3$ and so forth.

   \noindent {\bf CS: revision.}  When a manuscript is revised, its
   quality and novelty are improved. The improvement tends to be smaller
   when the manuscript has been revised several times. The quality of
   the revised manuscript $q'$ is computed as $h(q,0.1,k)$ where $k$ is
   the revision number and $h(a,b,c)$ is the following function:
   \begin{equation}
     h(a,b,c)=a+\frac{b}{c} (1-a) U[0,1]
   \end{equation}
   The manuscript can therefore improve up to 10\% after the first
   revision, but the subsequent improvements will be smaller and
   smaller. The same happens with the novelty. After the manuscript has
   been revised, the reviewers give a second estimate. Both the revision
   and the second estimate are instantaneous.

   \noindent {\bf CS: decision.} The probability of acceptance for the
   journal $X$ with parameters $\alpha_T^X, \ldots, \beta_N^T$ of the
   manuscript defined by the triplet $(t_e,q_e,n_e)$ is:
   \begin{equation}
     p(accept | X, t_e,q_e,n_e) = z(\alpha_T^X,\beta_T^X, t_e,0.1) \cdot
     \int_0^{q_e} B(\alpha_Q^X,\beta_Q^X,y') dy' \cdot \int_0^{n_e}
     B(\alpha_N^X,\beta_N^X,y'') dy''
   \end{equation}
   Once the probability of acceptance for a manuscript has been
   computed, the editor draws a random number from $U[0,1]$ and if the
   number is lower or equal than the probability, she accepts the
   article for publication, otherwise she rejects the manuscript. The
   decision for one manuscript happens the month after the reviews are
   completed. This acceptance function is used by each author to decide
   where to submit the manuscript: the author applies the function to
   her manuscript for all the journals, which then are ranked as
   explained above.
   
   \noindent {\bf AS: summary of algorithm.}  In the $AS$ a manuscript
   is created as in the $CS$. The following month, however, instead of
   being submitted to a journal, it is sent to the first pool. When the
   author submits a manuscript, she has to commit to choose three
   manuscript for review the next month (this is required in order to
   keep the minimum time before acceptance equal for $CS$ and
   $AS$). Once a manuscript has been reviewed, it is revised, the
   reviewers give a second opinion (instantaneous), and the manuscript
   is moved to the second pool. The following month, all the journals
   run the acceptance procedure on all the manuscripts in the second
   pool. If they are willing to accept any one of them, they bid on
   it. At the end of the month, authors choose for their manuscripts the
   journal with the highest impact among those that bid on it.
   
 no \bibliography{Bib_PR}

\section*{Acknowledgments} 
E. Baskerville, A.E. Budden, A.M. de Roos, E. Hackett, C. Lortie,
C. Melian, J. Parker and J. Ranganathan provided valuable suggestions
and comments. Part of this work was carried out when S.A.  was a
postdoctoral associate at the National Center for Ecological Analysis
and Synthesis, a center funded by National Science Foundation grant
DEB-0072909, and the University of California, Santa Barbara.

 %% FIGURES
 \begin{figure}
   \centering
   \includegraphics[height=0.99\textheight]{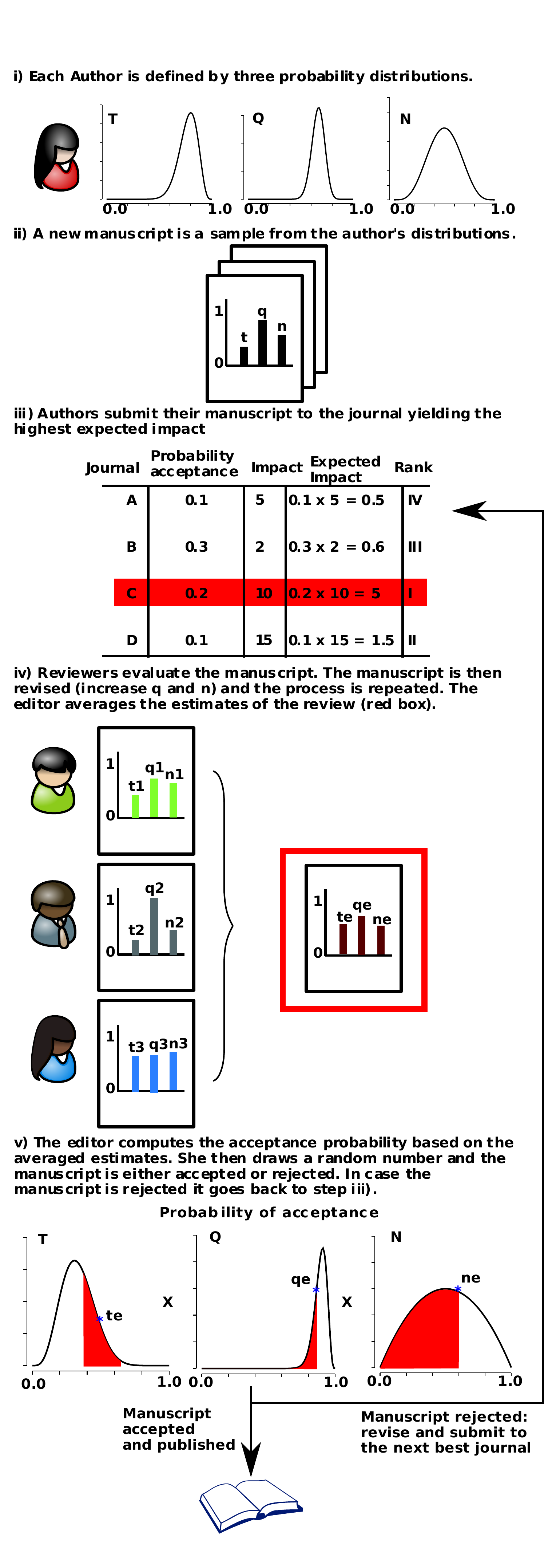}
   \caption{A flowchart for the Current System ($CS$) of peer
     review. }\label{fig:F1}
 \end{figure}
 
 \begin{figure}
   \centering
   \includegraphics[height=0.8\textheight]{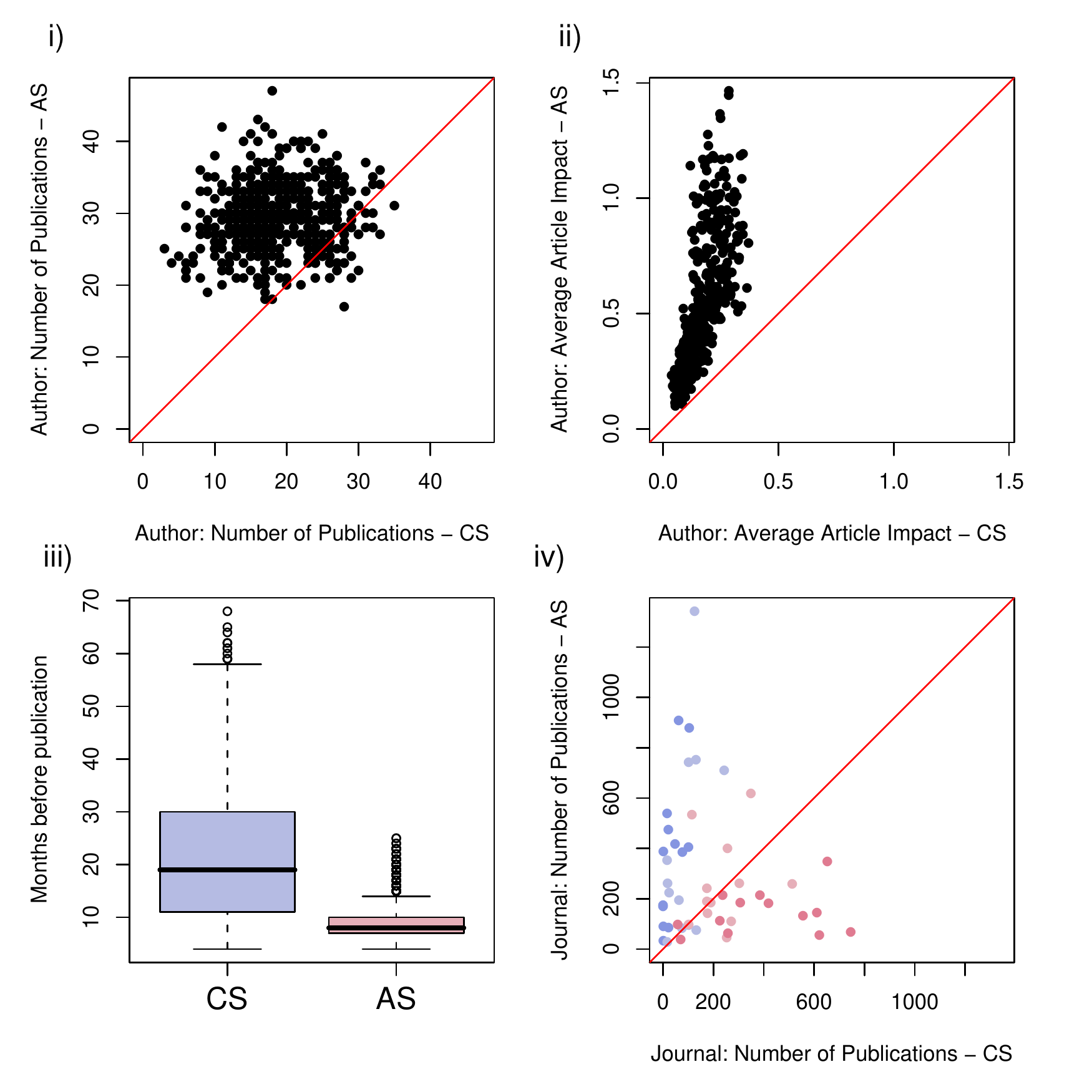}
   \caption{i) Comparison among the number of articles published by each
     author in the $CS$ ($x$ axis) and $AS$ ($y$ axis). ii) Average
     impact for each author in the $CS$ and $AS$. The impact is obtained
     by summing the impact of the journal in which each of the author's
     articles have been published and dividing by the number of
     publications. iii) Box-plot for the number of months before
     publication in the two settings. The thick black line represent the
     median. iv) Number of articles published by each journal in the two
     settings. The color indicates the impact of the journal: blue - top
     25\%, red - bottom 25\%. Lighter shadows indicate the quartiles
     around the mean.}\label{fig:F1}
 \end{figure}
 % \renewcommand{\figurename}{Figure S}
 % \setcounter{figure}{0}
 
 % \begin{figure}
 %   \centering
 %   \includegraphics[width=0.5\linewidth]{Results/S1.pdf}
 %   \caption{A simple food web composed by 5 species.  }\label{fig:S1}
 % \end{figure}
\end{document}